\documentclass[aps,twocolumn]{revtex4}
\pdfoutput=1

\usepackage{graphicx}
\usepackage{times}
\usepackage{amsmath}
\usepackage{amssymb}

\begin{document}

\title{Conformational selection in protein binding and function}

\author{Thomas R.\ Weikl and Fabian Paul\\[0.2cm]
\small Max Planck Institute of Colloids and Interfaces, Department of Theory and Bio-Systems \\[-0.1cm]
\small Science Park Golm, 14424 Potsdam, Germany}

\begin{abstract}
Protein binding and function often involves conformational changes. Advanced NMR experiments indicate that these conformational changes can occur in the absence of ligand molecules (or with bound ligands), and that the ligands may `select' protein conformations for binding (or unbinding). In this review, we argue that this conformational selection requires transition times for ligand binding and unbinding that are small compared to the dwell times of proteins in different conformations, which is plausible for small ligand molecules. Such a separation of timescales leads to a decoupling and temporal ordering of binding/unbinding events and conformational changes. We propose that conformational-selection and induced-change processes (such as induced fit) are two sides of the same coin, because the temporal ordering is reversed in binding and unbinding direction. Conformational-selection processes can be characterized by a {\em conformational excitation} that occurs {\em prior to} a binding or unbinding event, while induced-change processes exhibit a characteristic {\em conformational relaxation} that occurs {\em after} a binding or unbinding event. We discuss how the ordering of events can be determined from relaxation rates and effective on- and off-rates determined in mixing experiments, and from the conformational exchange rates measured in advanced NMR or single-molecule FRET experiments. For larger ligand molecules such as peptides, conformational changes and binding events can be intricately coupled and exhibit aspects of conformational-selection and induced-change processes in both binding and unbinding direction. 
\end{abstract}

\maketitle


%
\section*{Conformational changes of proteins in the native state}

In the last years, advances in single-molecule spectroscopy \cite{Min05,Michalet06,Smiley06,Kim13} and nuclear magnetic resonance (NMR)\cite{Palmer04,Mittermaier06,Boehr06b,Henzler07,Loria08,Lange08,Clore09} made it possible to detect and investigate higher-energy, excited-state conformations of folded proteins. The higher-energy conformations are in dynamic, thermally activated exchange with lowest-energy, ground-state conformations that correspond to the structures determined by x-ray crystallography or standard NMR methods. Structural investigations of such higher-energy conformations revealed that the thermally activated conformational changes are comparable to the conformational changes observed
during the binding or unbinding of ligands or during chemical reactions catalyzed by the proteins \cite{Eisenmesser05,Beach05,Boehr06a,Henzler07b,Lange08}, 
which demonstrated that these functional conformational changes  can occur in the absence of ligand or substrate molecules, or with bound ligands.

A theoretical basis for the conformational dynamics of proteins in the folded, native state is provided by energy-landscape perspectives, which have been first developed for protein folding \cite{Dill97,Bryngelson95,Dill85,Bryngelson87}. Key concepts are the {\em population shift} of protein conformations during binding or chemical reactions, and {\em conformational selection}, i.e.\ the view that pre-existing, higher-energy conformations of proteins can be `selected'  by ligands for binding \cite{Ma99}. Central aspects of these concepts already emerged in early models of protein allostery \cite{Monod65,Koshland66,Changeux11}.

\section*{Decoupling and temporal ordering of conformational changes and binding events for small binding transition times}

In general, both conformational changes and binding/unbinding events are thermally activated process that require the crossing of free-energy barriers. A characteristic feature of thermally activated processes is that the actual time for crossing the free-energy barrier, the `transition time', is significantly smaller than the dwell times in the states before and after the barrier crossing. In single-molecule experiments that probe conformational changes of proteins, the transition times are typically beyond experimental resolution, and the conformational changes appear as `sudden jumps' between two or more conformational states \cite{Schuler08,Chung12,Zoldak13,Woodside14,Kim13}. Similarly, thermally activated binding and unbinding events have typical transition times that are significantly smaller than the dwell times in the bound and unbound state. The transition time for binding is the time for crossing the energy barrier into the binding site from a position in direct proximity of the binding site. This binding transition time is independent of the molecular concentrations, in contrast to the dwell times in the unbound state, which are related to the binding rate. 

Let us consider now a protein $P$ with two dominant conformations 1 and 2 that can both bind to a ligand $L$. If this ligand is a small molecule, it seems reasonable to assume that the transition times for binding and unbinding are small compared to the dwell times in the conformational states of the protein with typical durations of submilliseconds to seconds. A direct consequence is that the binding and unbinding events of the ligand then are decoupled from the conformational transitions, because these binding and unbinding events occur during the dwell time the protein spends in either conformation 1 or conformation 2. This decoupling leads to a temporal ordering of binding/unbinding events and conformational changes: the binding event occurs either before or after the conformational change (see Fig.\ 1). 

Conformational changes and binding or catalytic events of larger ligands, in contrast, can be intricately coupled \cite{Csermely10}. The binding-induced folding of peptides, for example, involves a coupling between the binding and the conformational change of the peptides from unfolded to folded without clear temporal ordering of binding events and conformational transitions \cite{Sugase07,Bachmann11}.  As another example, the proline cis-trans-isomerization of peptide substrates that is catalyzed by the enzyme cyclophilin A exhibits a close coupling between the conformational change of the enzyme and the conformational isomerization of the peptide during the catalysis \cite{Eisenmesser02,Eisenmesser05}.  

\begin{figure}[t]
\begin{center}
\resizebox{0.6\columnwidth}{!}{\includegraphics{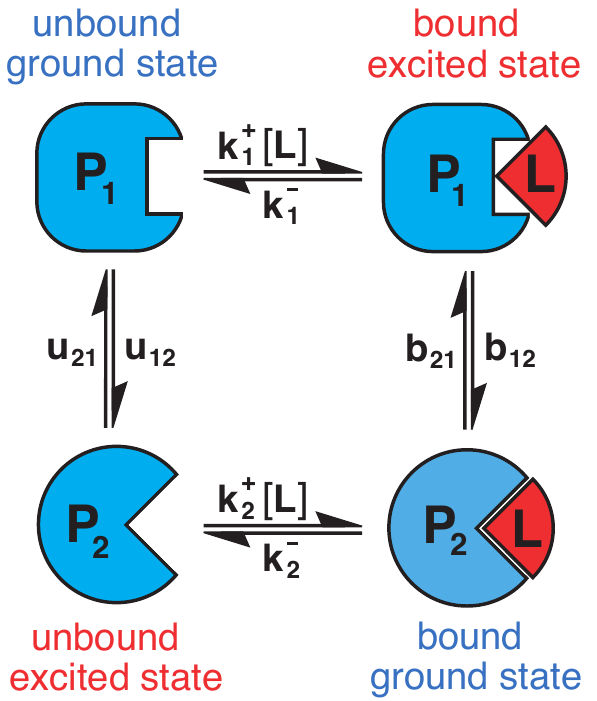}}
\end{center}
\caption{Four-state model of a protein with two conformations $P_1$ and $P_2$ that can both bind to a ligand $L$  \cite{Bosshard01,Sullivan08,Weikl09,Boehr09,Hammes09}. In the unbound state of the protein, the conformation $P_1$ is the lower-energy, ground-state conformation, while $P_2$ is the higher-energy, excited-state conformation.  In the bound state of the protein, $P_2L$ is the lower-energy ground state, and $P_1L$ is the higher-energy excited state. The bound ground state $P_2L$ is stabilized by more favorable interactions between the protein and the ligand, compared to the protein-ligand interactions in the state $P_1L$. In the unbound state of the protein, the rate $u_{12}$ for the transition from $P_1$ to $P_2$ is an excitation rate that is much smaller than the relaxation rate $u_{21}$ back into the ground state $P_1$. In the bound state, the rate $b_{12}$ is the relaxation rate in the ground state $P_2L$, and $b_{21}\ll b_{12}$ is the excitation rate. Under `pseudo-first order' conditions with a ligand concentration $[L]$ that greatly exceeds the protein concentration, the rates for the binding transitions in the two conformations are $k_1^{+}[L]$ and  $k_2^{+}[L]$.
}
\label{figure_4state_model}
\end{figure}
\section*{Conformational selection and induced changes are two sides of the same coin}

Along the lower pathway of Fig.\ 1 from the unbound ground state $P_1$ to the bound ground state $P_2L$, the conformational change occurs {\em prior to} the binding of the ligand. This binding mechanism has been termed `conformational selection' \cite{Ma99}, because the ligand appears to select conformation $P_2$ for binding. The conformational change along this pathway is a thermally activated transition from the unbound ground-state conformation $P_1$ to the excited-state conformation $P_2$.  Along the upper pathway of Fig.\ 1 from $P_1$ to $P_2L$, the conformational change of the protein occurs {\em after} the binding of the ligand to the unbound ground-state conformation $P_1$. This alternative binding mechanism has been termed `induced fit' \cite{Koshland58} because the conformational change along this pathway is apparently induced by the binding of the ligand. The conformational change along this pathway is a relaxation process from the bound excited state $P_1L$ into the bound ground state $P_2L$. Conformational selection and induced fit thus can be identified by the {\em temporal ordering} of the `chemical' binding step and the `physical' conformational change along the pathway. For induced-fit binding, the conformational change occurs after the binding step. For conformational-selection binding, the  conformational change occurs prior to the binding step.

\begin{figure}[b]
\begin{center}
\resizebox{0.9\columnwidth}{!}{\includegraphics{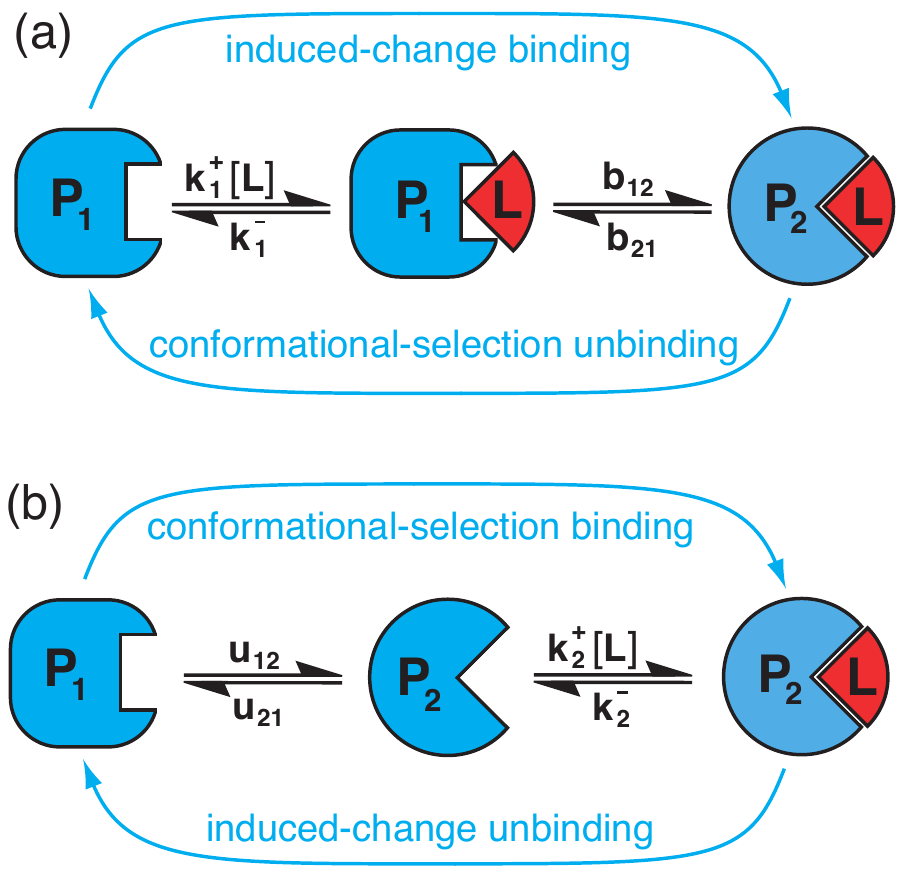}}
\end{center}
\caption{(a) Upper pathway of Fig.\ 1 from the unbound ground state $P_1$ to the bound ground state $P_2L$. Along this pathway, the conformational change occurs {\em after} the binding of the ligand molecule $L$, and is apparently `induced' by this binding event. In the reverse direction from $P_2L$ to $P_1$, the conformational change occurs {\em prior to} the unbinding of the ligand. The ligand thus appears to `select' the excited-state conformation $P_1L$ for unbinding.  (b) Lower pathway of Fig.\ 1 from the unbound ground state $P_1$ to the bound ground state $P_2L$. Along this pathway, the ligand binds {\em via} conformational selection, i.e.\ the conformational change occurs {\em prior to} ligand binding. In the reverse direction from $P_2L$ to $P_1$, the conformational change is induced by the unbinding of the ligand. The conformational transitions of the induced-change processes in (a) and (b) are relaxations into a ground state. The conformational changes of the conformational-selection processes, in contrast, are excitations.
}
\label{figure_3state_models}
\end{figure}
\begin{figure*}[t]
\begin{center}
\resizebox{1.25\columnwidth}{!}{\includegraphics{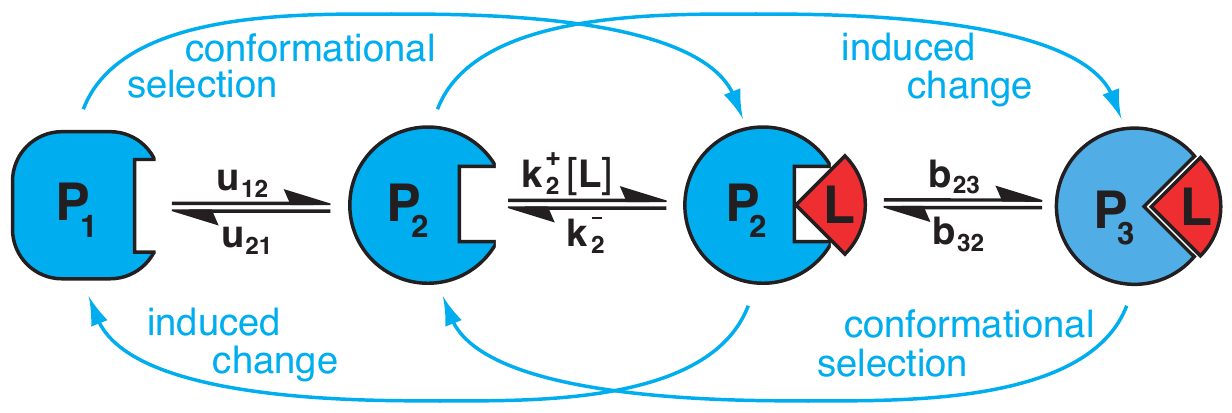}}
\end{center}
\caption{Four-state pathway on which a ligand $L$ binds in an intermediate conformation $P_2$ of a protein that is neither the unbound ground-state nor the bound ground-state conformation \cite{Wlodarski09,Weikl12,Vogt13}. This pathway exhibits sequential conformational-selection and induced-change processes in both binding and unbinding direction. The temporal ordering of the conformational changes and binding events of the pathway requires transition times for ligand binding and unbinding that are small compared to the dwell times in the states $P_2$ and $P_2L$ (see text). The effective on- and off-rate constants of the pathway are $k_\text{on} \simeq u_{12}k_{2}^{+}b_{23}/(u_{21} k_{2}^{-} + u_{21}b_{23} +  k_{2}^{+}[\text{L}]b_{23})$ and $k_\text{off} \simeq u_{21}k_{2}^{-}b_{32}/(u_{21} k_{2}^{-} + u_{21}b_{23} +  k_{2}^{+}[\text{L}]b_{23})$ \cite{Weikl12}.
}
\label{figure_sequential_model}
\end{figure*}

In thermal equilibrium, the flux along the binding direction of each of the two pathways, i.e.\ the number of binding events per unit time, is equal to the opposite flux in unbinding direction, because the principle of detailed balance prohibits any flux cycles in equilibrium. Fluxes along a cycle only exist in the presence of a non-equilibrium driving force. For example, non-equilibrium concentrations of substrate and product molecules drive the flux along the catalytic cycle of an enzyme \cite{Hill89}. 

Let us assume now that one of the two pathways of Fig.\ 1 is dominant, and that the flux along the other pathway is negligible. A direct consequence of the detailed-balance principle then is that the same pathway dominates in the reverse direction. This implies that the ordering of the `chemical' and `physical' substeps is reversed in the unbinding direction, compared to binding. In the reverse direction of the induced-fit binding pathway, the `physical' conformational change from the bound ground state $P_2L$ to the bound excited state $P_1L$ occurs {\em prior to} the `chemical' unbinding step of the ligand. The reverse of induced-fit binding thus is unbinding {\em via} conformational selection, because the ligand `selectively' unbinds from the excited-state conformation $P_1L$, after the conformational change (see Fig.\ 2(a)). In the reverse direction of the conformational-selection binding pathway, the conformational relaxation from the unbound excited state $P_2$ into the unbound ground state $P_1$ occurs after the unbinding step. This conformational relaxation is `induced' by the unbinding of the ligand (see Fig.\ 2(b)). 

Conformational selection and induced conformational changes thus can be seen as two sides of the some coin: a conformational-selection mechanism turns into an induced-change mechanism  when the overall direction of binding or unbinding is reversed. In this view, conformational selection is identified by a conformational change that occurs {\em prior to} a chemical step, which may be binding, unbinding, or also a catalytic step. This conformational change is a conformational excitation from a lowest-energy, ground-state conformation to a higher-energy conformation. Induced conformational changes, in contrast, are identified by a conformational change that occurs {\em after} a chemical step. These induced conformational changes are conformational relaxations from an excited-state to a ground-state conformation. The notion of `induced change' here generalizes the notion of `induced fit' to unbinding events and catalytic steps, i.e.\ to chemical steps beyond binding.  The temporal ordering of physical and chemical steps also helps to identify conformational selection and induced-change processes along more complex pathways such as the pathway shown in Fig.\ 3 along which a ligand binds to a protein confirmation that is different both from the unbound ground-state and the bound ground-state conformation.

\begin{figure*}[t]
\begin{center}
\resizebox{1.3\columnwidth}{!}{\includegraphics{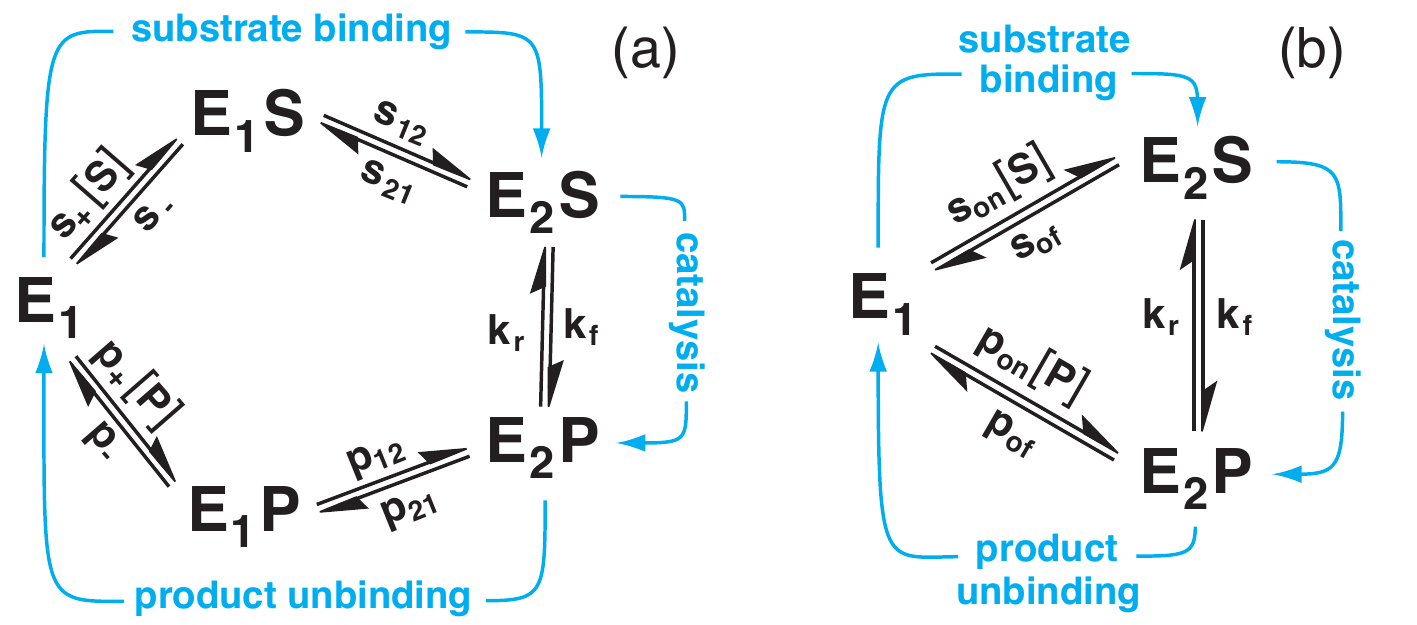}}
\end{center}
\caption{(a) Catalytic cycle of an enzyme with induced-change binding and conformational-selection unbinding of substrate and product molecules. Such a cycle has been suggested for the HIV-1 protease \cite{Furfine92,Gustchina90} and other enzymes \cite{Sullivan08} that change from an open conformation 1 to a closed conformation 2 during binding of substrate and product molecules and that have a binding site that is sterically blocked in the closed conformation. Along this cycle, $E_2S$ and $E_2P$ are the substrate-bound and product-bound ground states, and $E_1S$ and $E_1P$ are excited states.
(b) Effective 3-state catalytic cycle that is obtained from the full 5-state cycle for negligible product rebinding and for conformational excitation rates $s_{21}$ and $p_{21}$ that are much smaller than the relaxation rates $s_{12}$ and $p_{12}$ into the bound ground states. Along this 3-state cycle, the substrate binding and product unbinding steps include conformational changes of the enzyme and are described by effective rate constants $s_\text{on} = s_{12} s_{+}/(s_{12}+s_{-})$, $s_\text{of} = s_{21} s_{-}/(s_{12}+s_{-})$, and $p_\text{of} = p_{21} p_{-}/(p_{12}+p_{-})$ that depend on the conformational transition rates  and binding and unbinding rate constants of the full 5-state model (see text)  \cite{Weikl13}. 
}
\label{figure_catalytic_cycles}
\end{figure*}
\section*{Effective, overall rates of conformational-selection and induced-change processes}
\subsubsection*{Effective rates from relaxation experiments}

Binding processes are often investigated in {\em relaxation experiments} \cite{Fersht99,James03,Kim07,Kiefhaber12}. In such experiments, two solutions that contain, e.g., either proteins or ligands are initially mixed, and the relaxation into the binding equilibrium is reflected in a time-dependent fluorescence intensity of the mixed solutions. The experiments rely on a change of fluorescence intensity of the ligand or protein upon binding. The relaxation process is typically described as a single-exponential or multi-exponential process. The characteristic rate of a single-exponential process and the dominant, slowest rate of a multi-exponential are termed $k_\text{obs}$. For a multi-exponential process, the rate $k_\text{obs}$ governs the final relaxation into equilibrium, while the other, larger rates describe initial, faster relaxation processes. 

Under `pseudo-first order' conditions with a ligand concentration $[L]$ that greatly exceeds the protein concentration, the relaxation process of the two-step binding mechanisms in Fig.\ 2(a) and (b) is a bi-exponential process \cite{Weikl09,Weikl12,Vogt12}. The slower, dominant rate of this process has the general from
\begin{equation}
k_\text{obs} =  k_1 = \frac{1}{2}\left(\sigma - \sqrt{\sigma^2 - 4 \rho}\right) 
\label{kobs}
\end{equation}
while the faster rate is $k_2 = \frac{1}{2}\left(\sigma + \sqrt{\sigma^2 - 4\rho}\right)$. For the {\em induced-change binding} and {\em conformational-selection unbinding} pathway of Fig.\ 2(a), we have 
\begin{align}
& \sigma = k^{+}_{1}[\text{L}] + k^{-}_{1} + b_{12} + b_{21}  
\label{sigmaA} \\
& \rho = k^{+}_{1}[\text{L}](b_{12} + b_{21}) + k^{-}_{1}b_{21} 
\label{rhoA} 
\end{align}
For the {\em conformational-selection binding} and {\em induced-change unbinding} pathway of Fig.\ 2(a), we have 
\begin{align}
& \sigma = u_{12} + u_{21} + k^{+}_{2}[\text{L}] + k^{-}_{2} 
\label{sigmaB}  \\
& \rho =  u_{12} (k^{+}_{2}[\text{L}] + k^{-}_{2}) + u_{21} k^{-}_{2}
\label{rhoB}
\end{align}

An important special case of the relaxation kinetics is when the intermediate, excited state has a much lower equilibrium probability than the two terminal ground states, i.e.\ when the excited state $P_1L$ in Fig.\ 2(a) is much lower in probability than the ground states $P_1$ and $P_2L$, and when the excited state $P_2$ in Fig.\ 2(b) has a much lower probability than $P_1$ and $P_2L$ \footnote{This is the case for ligand concentrations $[L]$ with $k_{1}^{+}[L]\ll k_1^{-}$ and $k_{2}^{+}[L]\gg k_2^{-}$ at which the binding equilibrium in Fig.\ 2(a)  between $P_1$ and $P_1L$ is shifted towards $P_1$, and the binding  equilibrium in Fig.\ 2(b) between $P_2$ and $P_2L$ is shifted towards $P_2L$, respectively. By definition, the conformational equilibria between $P_1L$ and $P_2L$ in Fig.\ 2(a) and between $P_1$ and $P_2$ in Fig.\ 2(b) are always shifted towards the ground states $P_2L$ and $P_1$, respectively, which implies $u_{12}\ll u_{21}$ and $b_{12}\gg b_{21}$.}. 
The dominant relaxation rate for the 3-state mechanisms of Fig.\ 2(a) and Fig.\ 2(b) then can be written in the characteristic form 
\begin{equation}
k_\text{obs} \simeq k_\text{on}[L] + k_\text{off}
\end{equation}
of an {\em effective 2-state mechanism}
\begin{equation}
P_1 \;
\begin{matrix} 
\text{\footnotesize $k_\text{on}[L]$}\\[-0.05cm]
\text{\Large $\rightleftharpoons$}\\[-0.15cm]
\text{\footnotesize $k_\text{off}$} \\[0.05cm]
\end{matrix}
 \; P_2L
\end{equation}
For the {\em induced-change binding} and {\em conformational-selection unbinding} pathway of Fig.\ 2(a), the {\em effective on- and off-rates} are 
\begin{align}
& k_\text{on} = \frac{k_{1}^{+}b_{12}}{k_{1}^{-} + b_{12}} 
\label{kon_induced_change} \\
& k_\text{off} = \frac{k_{1}^{-}b_{21}}{k_{1}^{-} + b_{12}}
\label{koff_conformational_selection}
\end{align}
For the {\em conformational-selection binding} and {\em induced-change unbinding} pathway of Fig.\ 2(b), the effective rates are  
\begin{align}
& k_\text{on} = \frac{u_{12}k_{2}^{+}}{u_{21}+k_{2}^{+}[\text{L}]} 
\label{kon_conformational_selection} \\
& k_\text{off} = \frac{u_{21}k_{2}^{-}}{u_{21}+k_{2}^{+}[\text{L}]}
\label{koff_induced_change}
\end{align}
The numerator of these effective rates simply is the product of the rates for the two substeps in on- or off-direction, while the denominator is the sum of the two `outgoing rates' from the intermediate states $P_1L$ and $P_2$, respectively \cite{Weikl12}.

\subsubsection*{Effective rates on catalytic cycles}

A different route that leads to the same expressions for the effective rates $k_\text{on}$ and $k_\text{off}$ of conformational-selection and induced-change processes starts from the steady-state catalytic rates of enzymes. Fig.\ 4(a) illustrates the catalytic cycle of an enzyme that engages its substrate and product molecule {\em via} the induced-change binding and conformational-selection unbinding mechanism of Fig.\ 2(a). For negliglible product rebinding, and for conformational excitation rates $s_{21}$ and $p_{21}$ that are much smaller than the relaxation rates $s_{12}$ and $p_{12}$ into the substrate- and product-bound ground states $E_2S$ and $E_2P$, the steady-state catalytic rate along this cycle can be written in the Michaelis-Menten form
\begin{align}
k_\text{cat} = \frac{k_\text{max} [S]}{K_m + [S]}
\label{MichaelisMenten}
\end{align}
with 
\begin{align}
& k_\text{max} \simeq \frac{k_f p_\text{of}}{k_f + k_r + p_\text{of}} 
\label{kmax} \\
& K_m \simeq \frac{(k_f + s_\text{of})p_\text{of} + k_r s_\text{of}}{(k_f + k_r + p_\text{of}) s_\text{on}} 
\label{Km} 
\end{align}
and effective on- and off-rates $s_\text{on} = s_{12} s_{+}/(s_{12}+s_{-})$, $s_\text{of} = s_{21} s_{-}/(s_{12}+s_{-})$, and $p_\text{of} = p_{21} p_{-}/(p_{12}+p_{-})$ that are identical with Eqs.\ (\ref{kon_induced_change}) and (\ref{koff_conformational_selection}) \cite{Weikl13}. With these effective on- and off-rates, the 5-state model of Fig.\ 4(a) can be reduced to the effective 3-state model shown in Fig.\ 4(b). The Eqs.\ (\ref{MichaelisMenten}) to (\ref{Km}) are the expressions for the catalytic rate $k_\text{cat}$ of this classical 3-state model of catalysis.

\subsection*{How can we identify the temporal ordering of conformational changes and binding/unbinding events?} 

Advanced NMR experiments can indicate excited-state conformations in bound or unbound states of proteins that are required for the conformational-selection or induced-change binding and unbinding mechanisms of Fig.\ 2 \cite{Eisenmesser05,Boehr06a,Henzler07b,Lange08}. The existence of excited-state conformations is necessary for both mechanisms, but is not yet sufficient to identify a binding mechanism. If both the bound and unbound excited state in the four-state model of Fig.\ 1 exist, both mechanisms of  Fig.\ 2 may be possible. For the enzyme {\em E.\ coli} DHFR, for example, excited-state conformations have been found for all chemical states along the catalytic cycle \cite{Boehr06a}. Based on experimental data, several lines of reasoning have been suggested to argue for or prove a binding mechanism.

\subsubsection*{Rate-limiting conformational excitation}

Two lines of reasoning to identify binding/unbinding mechanisms of proteins involve a comparison of conformational transition rates to effective, overall rates for binding or unbinding. A first line of reasoning is based on an agreement of conformational excitation rates determined in advanced NMR experiments with overall, effective rates obtained from relaxation experiments \cite{Boehr06a}. Such an agreement is typical for conformational-selection processes with a rate-limiting conformational transition. In the case of conformational-selection binding, the effective on-rate (\ref{kon_conformational_selection}) is $k_\text{on}[L] \simeq u_{12}$ for $u_\text{21}\ll k_2^{+}[L]$ where $u_\text{12}$ is the conformational excitation rate in the unbound state (see Figs.\ 1 and 2). 
In the case of conformational-selection unbinding, the effective off-rate (\ref{koff_conformational_selection}) is $k_\text{off} \simeq b_{21}$ for $k_1^{-} \gg b_{12}$ where $b_{21}$ is the conformational excitation rate in the bound state. The effective on- and off-rates  (\ref{kon_induced_change}) and (\ref{koff_induced_change}) of induced-change binding and unbinding, in contrast, do not depend on the conformational excitation rates. 
An agreement between conformational excitation rates and effective rates therefore makes a conformational-selection mechanism plausible, but does not strictly rule out an alternative induced-change mechanism.

A second line of reasoning can be based on the fact that conformational excitation rates are an upper limit for the effective rates of conformational-selection processes. For conformational-selection binding, we have $k_\text{on}\le u_{12}$ for all possible values of the rate constants in Eq.\ (\ref{kon_conformational_selection}), i.e.\ the effective on-rate $k_\text{on}$ cannot be larger than the excitation rate $u_{12}$ for the conformational transition in the unbound state. For conformational-selection unbinding, we $k_\text{off}\le b_{21}$ where $b_{21}$ is the conformational excitation rate in the bound state (see Eq.\ \ref{koff_conformational_selection}). Effective off-rates determined from relaxation experiments that are larger than conformational excitation rates from advanced NMR experiments therefore rule out conformational-selection processes.  

\subsubsection*{Steric effects} 

Many proteins change from an open to a closed conformation during the binding of ligands. Steric effects can prohibit the entry or exit of ligands in closed protein conformations, e.g.\ if a lid closes over a ligand in such a conformation \cite{Sullivan08}. The proteins can then bind their ligands only {\em via} the induced-change binding mechanism of Fig.\ 2(a), because the proteins have to bind their ligands in the open conformation 1, prior to the conformational change to the closed conformation 2. Unbinding then has to occur {\em via} the conformational-selection unbinding mechanism of Fig.\ 2(a). 

Proteins that can only bind their ligands in an open conformation may still exhibit conformational changes in the unbound state between an open and a closed conformation \cite{Kim13}. However, these conformational changes are not `on pathway' regarding binding and illustrate that observed conformational changes in unbound states are not sufficient to identify a binding mechanism.

\subsubsection*{Relaxation kinetics}

The dominant, slowest relaxation rates $k_\text{obs}$ for the two binding mechanisms in Fig.\ 2 depend on the ligand concentration $[L]$ (see Eqs.\ (\ref{kobs}) to (\ref{rhoB})) \cite{Tummino08,Weikl09,Weikl12,Vogt12,Vogt13}. For the induced-change binding and conformational-selection unbinding mechanism of Fig.\ 2(a), the dominant relaxation rate $k_\text{obs}$ given in Eqs.\ (\ref{kobs}) to (\ref{rhoA})  always increases with the ligand concentration $[L]$. For the conformational-selection binding mechanism of Fig.\ 2(b), the relaxation rate $k_\text{obs}$ given in Eqs.\ (\ref{kobs}),  (\ref{sigmaB}), and (\ref{rhoB}) increases with $[L]$ for a fast conformational excitation rate $u_{12} > k_2^{-}$, but decreases with $[L]$ for a slow conformational excitation rate $u_{12} < k_2^{-}$ \cite{Vogt12}. A decrease of $k_\text{obs}$ with the ligand concentration $[L]$ observed in experiments thus indicates conformational-selection binding, since $k_\text{obs}$ always increases with $[L]$ in the case of induced-change binding. In contrast, an increase of $k_\text{obs}$ with the ligand concentration $[L]$ can both occur for conformational-selection binding and induced-change binding.

\begin{table*}
{\bf Table 1}: Effect of distal mutations on effective on- and off-rates $k_\text{on}$ and $k_\text{off}$
\label{table_mutations}
\begin{center}
\begin{tabular}{c|c|c}
 & ~~~ conformational-selection binding ~~~ & ~~~~~~~~ induced-change binding ~~~~~ \\[0.1cm]
\hline\\[-0.3cm]
 fast conformational relaxation ~ & 
$k_\text{on}^\prime/k_\text{on}= \exp(-\Delta\Delta G_u/RT)$   & $k_\text{on}^\prime/k_\text{on} = 1$   \\
 & $k_\text{off}^\prime/k_\text{off} = 1$ & $k_\text{off}^\prime/k_\text{off} = \exp(\Delta\Delta G_b/RT)$ \\[0.2cm]
\hline\\[-0.3cm]
 slow conformational relaxation ~ &
$ k_\text{on}^\prime/k_\text{on} = u_{12}^\prime/u_{12}$ & $k_\text{on}^\prime/k_\text{on} = b_{12}^\prime/b_{12}$  \\ 
 &$ k_\text{off}^\prime/k_\text{off} = u_{21}^\prime/u_{21}$ & $k_\text{off}^\prime/k_\text{off} = b_{21}^\prime/b_{21}$ \\
\end{tabular}
\end{center}
\end{table*}

\subsubsection*{Effect of distal mutations}

Mutations distal to the binding site that mainly affect the conformational exchange, but not the binding kinetics in different protein conformations, can inform on the binding mechanism. Such distal mutations have also been termed allosteric mutations, or non-active site mutations. Of particular interest is how such mutations change the effective, overall on- or off-rates of the two binding/unbinding mechanisms of Fig.\ 2. To analyze these changes, we distinguish two cases:

{\em Case 1 -- fast conformational relaxation}: The effective on- and off-rate constants (\ref{kon_induced_change}) and (\ref{koff_conformational_selection}) for the {\em induced-change binding} and {\em conformational-selection unbinding} pathway of Fig.\ 2(a) simplify to
\begin{align}
k_\text{on} \simeq k_1^{+} \text{~~and~~} k_\text{off} \simeq k_1^{-}b_{21}/ b_{12}
\end{align}
for a large conformational relaxation rate $b_{12}\gg k_1^{-}$ of the transition into the bound ground-state. Mutations that only affect the conformational transition rates  $b_{21}$ and  $b_{12}$, but not the binding rate constants $k_1^{+}$ and $k_1^{-}$ therefore do not change the effective on-rate constant $k_\text{on}$ for induced-change binding. The effect of such mutations on the off-rate constant $k_\text{off}$ can be quantified by the change of the free-energy difference $\Delta G_b = G(P_2L) - G(P_1L) = - RT\ln(b_{12}/b_{21})$ between the two bound conformations where $R$ is the gas constant and $T$ is temperature (see Table 1).  For the {\em conformational-selection binding} and {\em induced-change unbinding} pathway of Fig.\ 2(b), the effective on- and off-rate constants (\ref{kon_conformational_selection}) and (\ref{koff_induced_change}) are
\begin{align}
k_\text{on} \simeq u_{12} k_2^{+}/ u_{21}   \text{~~and~~} k_\text{off} \simeq k_2^{-}
\end{align}
in the case of a large conformational relaxation rate $u_{21} \gg k_2^{+}[L]$. Distal mutations that do not change the binding und unbinding rates $k_2^{+}$ and $k_2^{-}$ therefore do not change the effective off-rate $k_\text{off}$, but can change $k_\text{on}$ by shifting the conformational free-energy difference $\Delta G_u = G(P_2) - G(P_1) = - RT\ln( u_{12}/u_{21})$  in the unbound state. For both pathways, distal mutations thus change the effective rate of the conformational-selection process, but not the effective rate of the induced-change process in reverse direction. 

{\em Case 2 -- slow conformational relaxation}: For a small conformational relaxation rate $b_{12}\ll k_1^{-}$, the effective on- and off-rate constants (\ref{kon_induced_change}) and (\ref{koff_conformational_selection}) for the {\em induced-change binding} and {\em conformational-selection unbinding} pathway of Fig.\ 2(a) are
\begin{align}
k_\text{on} \simeq b_{12}k_1^{+}/k_1^{-} \text{~~and~~} k_\text{off} \simeq b_{21}
\end{align}
In this case, the ratio $k_\text{on}^\prime/k_\text{on}$ of the effective on-rate constants for a distal mutant and the wildtype is equal to ratio 
$b_{12}^\prime/b_{12}$ of the conformational {\em relaxation} rates, whereas the ratio $k_\text{off}^\prime/k_\text{off}$  is equal to the ratio $b_{21}^\prime/b_{21}$ of the conformational excitation rates. For the {\em conformational-selection binding} and {\em induced-change unbinding} pathway of Fig.\ 2(b), the effective rates simplify to
\begin{align}
k_\text{on}[L] \simeq u_{12}  \text{~~and~~} k_\text{off} \simeq u_{21} k_2^{-}/k_2^{+}[L]
\end{align}
for a small conformational relaxation rate $u_{21}\ll k_2^{+}[L]$ (see Eqs.\ (\ref{kon_conformational_selection}) and (\ref{koff_induced_change})). For this pathway, the ratio $k_\text{on}^\prime/k_\text{on}$ of the effective on-rate constants for a distal mutant and the wildtype is equal to the ratio $u_{12}^\prime/u_{12}$ of the {\em excitation} rates, and the ratio $k_\text{off}^\prime/k_\text{off}$  is equal to the ratio $b_{21}^\prime/b_{21}$ of the conformational {\em relaxation} rates. 

The remaining question for our analysis now is: does the mutation mainly affect the excitation rate or the relaxation rate of the conformational transitions, or both rates? Two scenarios can be distinguished: First, suppose the transition state for the conformational transition is close in free energy to the excited state, relative to the free-energy difference between excited state and ground state. According to the Hammond-Leffler postulate \cite{Hammond55,Leffler53}, the structure of the transition state then resembles the structure of the excited state, which implies that the mutations have a similar effect on the two states. The mutations then should not affect the free-energy difference between the excited state and the transition state, and should not change the relaxation rate for the transition to the ground state, which depends on this free-energy difference. The mutations mainly affect the excitation rates of the conformational transitions and, thus, the effective rates for the conformational selection processes of the two pathways, which depend on the excitation rates.  In the second scenario, the free-energy difference between the transition state and the excited state is large (or at least not small) compared to the free energy difference between the excited state and the ground state. Distal mutations then should in general affect both the excitation rate and the relaxation rate and, thus, both the effective on- and off-rates of the two pathways.

In summary, distal mutations that change effective on-rates but not effective off-rates therefore point towards the conformational-selection binding and induced-change unbinding pathway of Fig.\ 2(b). The effective off-rates are not changed either because of the fast conformational relaxation in case 1 above, or because of a transition state that is close in free energy to the excited state in case 2 of a slow conformational relaxation. In contrast, distal mutations that change effective off-rates but not effective on-rates point towards the induced-change binding and conformational-selection unbinding pathway of Fig.\ 2(a).  For example, distal mutations in the clyclic nucleotide binding domain (CNBD) of a nucleotide-gated channel from {\em Mesorhizobium loti} have been found to affect mainly the overall off-rate, which indicates induced-change binding and conformational-selection unbinding  of this domain \cite{Peuker13}. Distal mutations that change both the effective on-rates and off-rates point towards a slow conformational relaxation with a transition state for the conformational transition that is significantly higher in free energy than the excited state, but do not allow to identify a binding mechanism.

\subsubsection*{Single-molecule FRET}

Sufficiently large conformational changes of proteins can be observed by single-molecule fluorescence resonance energy transfer (FRET) experiments. In these experiments, an acceptor and a donor molecule are attached to a protein, and transitions between, e.g., an open conformation 1 and closed conformation 2 are reflected by sudden changes in the distance-dependent transfer efficiency of the two molecules. The experiments are conducted in equilibrium. For the {\em induced-change binding} and {\em conformational-selection unbinding} pathway of Fig.\ 2(a), the `closing rate' $k_{12}$ and `opening rate' $k_{21}$ observed in such experiments are:
\begin{align}
& k_{12} = \frac{p(P_1L)}{p(P_1) + p(P_1L)} b_{12}= \frac{k_1^{+}[L]}{k_1^{-} + k_1^{+}[L]} b_{12} 
\label{k12a}\\ 
& k_{21} = b_{21}
\label{k21a}
\end{align}
Here, $p(P_1)$ and $p(P_1L)$ are the equilibrium probabilities of the states $P_1$ and $P_1L$. The closing rate $k_{12}$ is the product of (i) the relative equilibrium probability $p(P_1L)/(p(P_1) + p(P_1L))$ of the state $P_1L$ among the two states $P_1$ and $P_1L$ with conformation 1 and (ii) the rate $b_{12}$ for the conformational transition from $P_1L$ to $P_2L$. The opening rate $k_{21}$ simply is equal to the rate $b_{21}$ for the reverse conformational transition. For small ligand concentrations $[L]$ with $k_1^{+}[L]\ll  k_1^{-}$, the opening rate $k_{12}$ of Eq.\ (\ref{k12a}) increases linearly with $[L]$. The closing rate $k_{21}$ of this pathway, in contrast, is independent of the ligand concentration $[L]$.

For the {\em conformational-selection binding} and {\em induced-change unbinding} pathway of Fig.\ 2(b), the `closing rate' $k_{12}$ and `opening rate' $k_{21}$ observed in single-molecule FRET experiments are
\begin{align}
& k_{12} = u_{12}\\
& k_{21} = \frac{p(P_2)}{p(P_2) + p(P_2L)} u_{21}= \frac{k_2^{-}}{k_2^{-} + k_2^{+}[L]} u_{21} 
\end{align}
where $p(P_2)$ and $p(P_2L)$ are the equilibrium probabilities of the states $P_2$ and $P_2L$. For this pathway, the opening rate $k_{21}$ decreases with increasing ligand concentration $[L]$, while the closing rate $k_{12}$ is independent of $[L]$. The dependence of the opening and closing rates on the ligand concentration $[L]$ thus is different for the two pathways of Fig.\ 2 and can help to identify a binding and unbinding mechanism. 
For example, single-molecule FRET experiments of the maltose-binding protein reveal an opening rate that increases linearly with the ligand concentration $[L]$ for small $[L]$ and a closing rate that is nearly independent of $[L]$ \cite{Kim13,Seo14}, which is characteristic for the induced-change binding and conformational-selection unbinding pathway of Fig.\ 2(a). However, the analysis of these experiments is complicated by the fact that the maltose-binding domain exhibits conformational changes both in the bound and unbound state \cite{Kim13}, which requires a more detailed analysis based on four-state models discussed in the Appendix.  

\section*{Conclusions and outlook}

In this review, we have focused on simple models for the conformational changes of proteins during binding and unbinding. In these models, the binding/unbinding events and conformational changes of the proteins are decoupled and have a clear temporal order. We have argued that this temporal order requires transition times for binding and unbinding that are small compared to the dwell times of the proteins in different conformations, which is plausible for small ligands, and have reviewed several lines of reasoning to extract the temporal order of binding/unbinding events and conformational changes based on data from relaxation experiments, advanced NMR experiments, or single-molecule FRET experiments. 

If proteins bind to larger ligand molecules, e.g.\ to peptides, other proteins, or DNA or RNA molecules, conformational changes and binding events can be intricately coupled \cite{Csermely10,Sugase07,Bachmann11}.  A promising approach to model such an intricate coupling is Markov state modeling of molecular dynamics simulations \cite{Bowman14,Noe07,Chodera07,Buchete08,Bowman09,Senne12}. 
Such Markov state modeling has been previously applied to obtain pathways for the conformational changes of proteins \cite{Noe09,Voelz10,Silva11,Sadiq12}. Other methods to explore rare conformational transitions of proteins with atomistic molecular dynamic simulations identify reaction coordinates or collective variables for such transitions \cite{Elber11,Piana07,Juraszek08,Abrams10,Kirmizialtin12}. 
Conformational changes  of proteins have also been investigated in molecular dynamics simulations with coarse-grained models \cite{Turjanski08,Hyeon09,Tehver10} and by normal mode analysis of elastic network models \cite{Bahar10,Alexandrov05,Dobbins08,Meireles11,Bastolla14}.



\subsection*{Acknowledgements}

The authors would like to thank David Boehr, Bahram Hemmateenejad, and Frank No{\'e} for enlightening collaborations on the topic of this review.

\begin{appendix}

\section*{Appendix: Interpretation of FRET experiments}

\subsection*{Induced-change binding}

Single-molecule FRET experiments of the maltose binding domain have been interpreted with a four-state model that includes the induced-change binding pathway of Fig.\ 2(a) and an additional, `off-pathway' conformational change in the unbound state \cite{Kim13}. This four-state model corresponds to the model shown in Fig.\ 1 with zero binding and unbinding rates $k_2^{+} = k_2^{-} = 0$ in the closed conformation 2. 

The equilibrium probabilities $p$ of the four states of this model follow from the equations
\begin{align}
& p(P_1) + p(P_2) + p(P_1L) + p(P_2L) = 1 
\label{p_normalization}\\
& p(P_1) u_{12} - p(P_2)  u_{21} = 0  \\ 
& p(P_2)u_{21} + p(P_1L) k_1^{-} - p(P_1) \left(u_{12} + k_1^{+}[L]\right) = 0 \\ 
& p(P_2L) b_{21} + p(P_1) k_1^{+}[L] - p(P_1L) (b_{12} + k_1^{-}) = 0 \\ 
& p(P_1L) b_{12} - p(P_2L) b_{21} = 0
\end{align}
Eq.\ (\ref{p_normalization}) simply states that the probabilities of all four states sum up to 1, while the remaining other four equations are the rate equations of the four states in equilibrium. The right-hand sides of these equations are zero because the influx and outflux of a state are oppositely equal in equilibrium. Of particular interest here are the relative probabitities $p_r$ of the two states $P_1$ and $P_1L$ with conformation 1
\begin{align}
& p_r(P_1) = \frac{p(P_1)}{p(P_1) + p(P_1L)} = \frac{k_1^{-}}{k_1^{-} + k_1^{+}[L]} \\
& p_r(P_1L) = \frac{p(P_1L)}{p(P_1) + p(P_1L)} = \frac{k_1^{+}[L]}{k_1^{-} + k_1^{+}[L]}
\end{align}
and the relative probabilities of the two states $P_2$ and $P_2L$ with conformation 2 
\begin{align}
& p_r(P_2) = \frac{p(P_2)}{p(P_2) + p(P_2L)} 
= \frac{b_{21}k_1^{-}u_{12}}{b_{21}k_1^{-}u_{12} + b_{12}k_1^{+}[L]u_{21}}\\
& p_r(P_2L) = \frac{p(P_2L)}{p(P_2) + p(P_2L)} 
= \frac{b_{12}k_1^{+}[L]u_{21}}{b_{21}k_1^{-}u_{12} + b_{12}k_1^{+}[L]u_{21}}\
\end{align}

The `closing' transition from conformation 1 to 2 and the reverse `opening' can either occur in the bound or unbound state of this model. Since the FRET experiments cannot distinguish between the bound and unbound state, the closing rate $k_{12}$ and opening rate $k_{21}$ observed in these experiments are
\begin{align}
& k_{12} = p_r(P_1)u_{12} + p_r(P_1L)b_{12} 
= \frac{k_1^{-}u_{12} + k_1^{+}[L] b_{12}}{k_1^{-} + k_1^{+}[L]} \\
& k_{21} = p_r(P_2)u_{21} + p_r(P_2L)b_{21} 
= \frac{\left(k_1^{-}u_{12} + k_1^{+}[L]b_{12}\right)b_{21} u_{21}}{b_{21}k_1^{-}u_{12} + b_{12}k_1^{+}[L]u_{21}}
\end{align}
An analysis of the sign of the derivatives of $k_{12}$ and $k_{21}$ with respect to the ligand concentration $[L]$ reveals that the closing rate  $k_{12}$ is monotonously increasing with $[L]$ for $b_{12} > u_{12}$, which is plausible because $b_{12}$ is a conformational relaxation rate, while $u_{12}$ is an excitation rate. The opening rate $k_{21}$, in contrast, is monotonously decreasing for $b_{21} < u_{21}$, which is plausible because $u_{21}$ is a conformational relaxation rate, while $b_{21}$ is an excitation rate.

\subsection*{Conformational-selection binding}

As an alternative model, we consider here the four-state model of Fig.\ 1 with zero and unbinding rates $k_1^{+} = k_1^{-} = 0$ in conformation 1. This alternative model combines conformational-selection binding as in Fig.\ 2(b) with an additional `off-pathway' conformational change in the bound state of the protein. Analogous to the calculations in the previous section, we now obtain the closing and opening rates
\begin{align}
& k_{12} = \frac{\left(k_2^{-}u_{21} + b_{21}k_2^{+}[L]\right)u_{12}b_{12}}{u_{21}k_2^{-}b_{12} + u_{12}k_2^{+}[L] b_{21}}\\
&k_{21} = \frac{k_2^{-}u_{21} + k_2^{+}[L] b_{21}}{k_2^{-} + k_2^{+}[L]}
\end{align}
Also in this model, the closing rate $k_{12}$ increases monotonously with the ligand concentration $[L]$  for $b_{12}> u_{12}$, and the opening rate $k_{21}$ decreases monotonously with $[L]$ for $b_{21} < u_{21}$, which is again plausible because $b_{12}$ and $u_{21}$ are the conformational relaxation rates in the bound and unbound state.  The dependence of the opening and closing rates on the ligand concentration $[L]$ thus is qualitatively similar in both four-state models considered in this appendix. An identification of the dominant binding mechanism therefore requires an analysis that goes beyond qualitative features of the functions $k_{12}[L]$ and $k_{21}[L]$.

\end{appendix}

\end{document}